\documentclass[sigconf,authorversion,nonacm]{acmart} 

\AtBeginDocument{%
  }

\newcommand{\hide}[1]{}

\newcommand{\srijan}[1]{\textcolor{red}{[Srijan: #1]}}

\usepackage{enumitem}
\usepackage{graphicx}
\usepackage{hyperref}  
\usepackage{natbib}
\usepackage{url}
\usepackage{epstopdf}
\usepackage{graphicx}
\usepackage{amsmath}
\usepackage{booktabs}
\usepackage{algorithm}
\usepackage[noend]{algorithmic}
\usepackage{comment}
\usepackage{pifont}
\usepackage{multirow}
\usepackage{mathtools}
\usepackage{balance}
\usepackage{color,xspace}
\usepackage{hyperref}
\usepackage{subfig}
\usepackage{caption}
\usepackage{amsthm}
\usepackage{soul}
\usepackage{bm}

\newcommand{\cmark}{\ding{51}}%
\begin{document}

\title{Representation Learning in Continuous-Time Dynamic Signed Networks}

\author{Kartik Sharma}
\authornote{MR and KS contributed equally to this work.}
\affiliation{%
  \institution{Georgia Institute of Technology}
}
\email{ksartik@gatech.edu}

\author{Mohit Raghavendra}
\authornote{This work was done during MR's internship at Georgia Tech}
\authornotemark[1]
\affiliation{%
  \institution{NIT Karnataka}
}
\email{mohithmitra@gmail.com}


\author{Yeon-Chang Lee}
\affiliation{%
  \institution{Georgia Institute of Technology}
}
\email{yeonchang@gatech.edu}

\author{Anand Kumar M}
\affiliation{%
  \institution{NIT Karnataka}
}
\email{m_anandkumar@nitk.edu.in}

\author{Srijan Kumar}
\affiliation{%
  \institution{Georgia Institute of Technology}
}
\email{srijan@gatech.edu}

%
%




\begin{abstract}
Signed networks allow us to model conflicting relationships and interactions, such as friend/enemy and support/oppose. 
These signed interactions happen in real time.
Modeling such dynamics of signed networks is crucial to understanding the evolution of polarization in the network and enabling effective prediction of the signed structure (i.e., link signs and signed weights) in the future. 
However, existing works have modeled either (static) signed networks or dynamic (unsigned) networks but not dynamic signed networks. 
Since both sign and dynamics inform the graph structure in different ways, it is non-trivial to model how to combine the two features. 
In this work, we propose a new Graph Neural Network (GNN)-based approach to model dynamic signed networks, named \textbf{\name}: \textbf{\ul{S}}igned link's \textbf{\ul{E}}volution using \textbf{\ul{M}}emory modules and \textbf{\ul{B}}alanced \textbf{\ul{A}}ggregation.
Here, the idea is to incorporate the signs of temporal interactions using separate modules guided by balance theory and to evolve the embeddings from a higher-order neighborhood.
Experiments on $4$ real-world datasets and $4$ different tasks demonstrate that \name consistently and significantly outperforms the baselines by up to $80\%$ on the tasks of predicting signs of future links while matching the state-of-the-art performance on predicting existence of these links in the future. We find that this improvement is due specifically to superior performance of \name on the minority negative class.

\end{abstract}%




\newcommand{\argmax}[1]{\underset{#1}{\operatorname{arg}\,\operatorname{max}}\;}
\newcommand{\argmin}[1]{\underset{#1}{\operatorname{arg}\,\operatorname{min}}\;}

\definecolor{applegreen}{rgb}{0.55, 0.71, 0.0}

\newcommand{\ks}[1]{\textcolor{applegreen}{KS: #1}}
\newcommand{\mr}[1]{\textcolor{blue}{MR: #1}}
\newcommand{\sk}[1]{\textcolor{blue}{SK: #1}}
\newcommand{\res}[1]{\textcolor{orange}{RES: #1}}

\newcommand{\gnn}{\textsc{GNN}\xspace}
\newcommand{\gnns}{\textsc{GNN}s\xspace}
\newcommand{\mlp}{\textsc{MLP}\xspace}
\newcommand{\gcn}{\textsc{GCN}\xspace}
\newcommand{\sage}{\textsc{SAGE}\xspace}
\newcommand{\sgcn}{\textsc{SGCN}\xspace}
\newcommand{\sigat}{\textsc{SiGAT}\xspace}
\newcommand{\lstm}{\textsc{LSTM}\xspace}
\newcommand{\sgclstm}{\textsc{SGC-LSTM}\xspace}
\newcommand{\tgn}{\textsc{TGN}\xspace}
\newcommand{\rwlstm}{\textsc{RW-LSTM}\xspace}
\newcommand{\name}{\textsc{SEMBA}\xspace}
\newcommand{\fname}{\textsc{\textbf{S}}igned links \textsc{\textbf{E}}volution using \textsc{\textbf{M}}emory modules and \textsc{\textbf{B}}alanced \textsc{\textbf{A}}ggregation\xspace}
\newcommand{\prop}{\textsc{Prop}\xspace}
\newcommand{\smem}{\textsc{SMem}\xspace}
\newcommand{\ba}{\textsc{BA}\xspace}

\newcommand{\btcotc}{\textsc{BTC-Otc}\xspace}
\newcommand{\btcalpha}{\textsc{BTC-Alpha}\xspace}
\newcommand{\bitcoinotc}{\textsc{Bitcoin-OTC}\xspace}
\newcommand{\bitcoinalpha}{\textsc{Bitcoin-Alpha}\xspace}
\newcommand{\epinions}{\textsc{Epinions}\xspace}
\newcommand{\wikirfa}{\textsc{WikiRfA}\xspace}
\newcommand{\sota}{\textsc{SoTA}\xspace}

\newcommand{\GG}{\mathcal{G}\xspace}
\newcommand{\GE}{\mathcal{E}\xspace}
\newcommand{\GEp}{\mathcal{E}^+\xspace}
\newcommand{\GEn}{\mathcal{E}^-\xspace}
\newcommand{\GV}{\mathcal{V}\xspace}
\newcommand{\EE}{\mathbb{E}\xspace}
\newcommand{\ev}{\texttt{ev}\xspace}

\newcommand{\msg}{\texttt{msg}\xspace}
\newcommand{\agg}{\texttt{agg}\xspace}
\newcommand{\att}{\texttt{att}\xspace}
\newcommand{\mem}{\texttt{mem}\xspace}
\newcommand{\emb}{\texttt{emb}\xspace}
\newcommand{\Nbrs}{\mathcal{N}\xspace}
\newcommand{\mbf}{\mathbf{m}\xspace}
\newcommand{\agmbf}{\mathbf{M}\xspace}
\newcommand{\sbf}{\mathbf{s}\xspace}
\newcommand{\ebf}{\mathbf{e}\xspace}
\newcommand{\zbf}{\mathbf{z}\xspace}
\newcommand{\xbf}{\mathbf{x}\xspace}
\newcommand{\Wbf}{\mathbf{W}\xspace}

\newcommand{\relu}{\texttt{ReLU}\xspace}

\newcommand{\fone}{\texttt{F1}\xspace}
\newcommand{\auc}{\texttt{AUROC}\xspace}
\newcommand{\aucwt}{\texttt{AUROC-wt}\xspace}
\newcommand{\fonemac}{\texttt{F1mac}\xspace}
\newcommand{\fonemic}{\texttt{F1mic}\xspace}
\newcommand{\acc}{\texttt{Acc}\xspace}
\newcommand{\fonewt}{\texttt{F1wt}\xspace}
\newcommand{\rmse}{\texttt{RMSE}\xspace}
\newcommand{\rsq}{$\texttt{R}^2$\xspace}
\newcommand{\kldiv}{\texttt{KL-div}\xspace}

\newcommand{\oom}{-\xspace}

\newcommand{\thetabf}{\mathbf{\Theta}\xspace}
\newcommand{\Ls}{\mathcal{L}\xspace}

\newtheorem{prob}{\textbf{Problem}}

\theoremstyle{plain} 
\newcommand{\thistheoremname}{}
\newenvironment{probl}[1]
  {\renewcommand{\thistheoremname}{#1}%
   \begin{genericthm}}
  {\end{genericthm}}



\maketitle

%
%


\section{Introduction}\label{sec:intro}

\begin{figure}[t]
    \centering
    \includegraphics[scale=0.8]{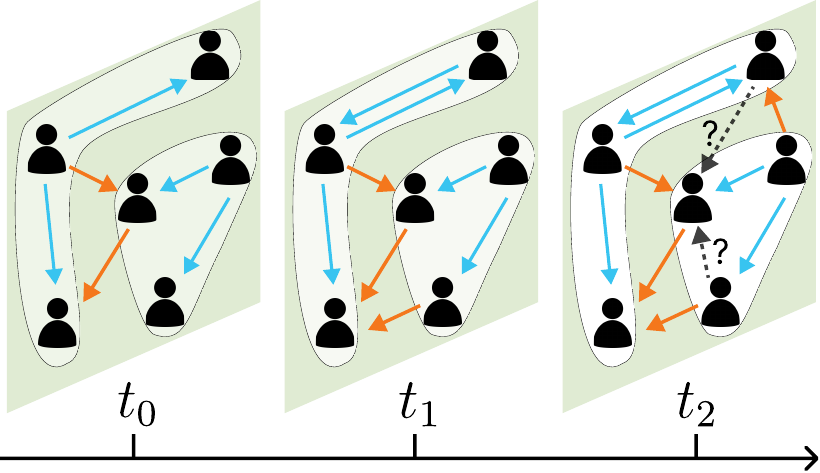}
    \vspace{-0.3cm}
    \caption{Visualizing Dynamic Signed Networks.}
    \label{fig:vis}
    \vspace{-0.5cm}
\end{figure}

\noindent\textbf{Background.} 
Social interactions are often signed and dynamic, representing the evolving nature of trust/distrust, support/oppose, agreement/disagreement, or positive/negative sentiments among people. 
Signed networks have been used to measure polarization in social networks \cite{bonchi2019discovering,alsinet2021measuring}, performance in financial networks \cite{askarisichani2019structural}, and predicting international conflicts \cite{doreian2019structural}. 
Dynamic signed networks model conflicting relationships that evolve over time, as shown in Figure~\ref{fig:vis}.
For instance, online social media discussions involve people supporting/opposing a claim, which continuously evolve as new users and new replies are made in real time. 
Dynamic signed networks have several societal applications such as predicting veracity of posts~\cite{shu2017fake}, toxicity in conversations~\cite{saveski2021structure}, conflict across communities~\cite{kumar2018community}, and analyzing polarization in debates~\cite{alsinet2021measuring,primario2017measuring,lai2018stance,lai2019stance}. 
Thus, it is important to create predictive models for dynamic signed networks 
so that one can devise intervention strategies against future unwanted behavior, maintaining a healthy online environment~\cite{walter2021evaluating,peleteiro2014exploring,axelrod2021preventing}. 

\hide{
The distribution of signs in these networks is often skewed to favor one of the signs (mostly, positive sign) and thus, the networks are often imbalanced in these signs. 
Due to this imbalance, making predictions from the raw data becomes a difficult task. In the literature, domain knowledge such as balance theory \cite{cartwright1956structural} is included in the learning system to effectively encode signed links.}
Structural balance theory models the organization of signed networks~\cite{cartwright1956structural}. It states that signed relationships tend to form balanced triads such that  ``a friend of my friend is my friend,'' ``a friend of my enemy is my enemy,'' ``an enemy of my friend is my enemy,'' and ``an enemy of my enemy is my friend.'' 
Several Signed Graph Neural Networks (GNNs) have been proposed to learn effective representations of static signed graphs, while incorporating balance theory~\cite{derr2018signed,huang2019signed,huang2021pole,LiuZ0ZCL021,ShuDC0ZXS21,HuangSHC21,LiTZC20}.

\hide{
With the advent of graph neural networks (GNN) \cite{kipf2016semi,velivckovic2017graph,wu2020comprehensive}, many researchers have extended existing GNNs for signed networks, thus developing signed GNNs \cite{derr2018signed,huang2019signed,huang2021pole,LiuZ0ZCL021,ShuDC0ZXS21,HuangSHC21,LiTZC20}. 
In a nutshell, they designed novel aggregation modules, which enable the incorporation of the balance theory into the GNN mechanism. 
For example, \sgcn \cite{derr2018signed} separately aggregates information coming from a balanced and an unbalanced path, i.e., from paths with an even and odd number of negative links, respectively. 
Literature has shown that such signed \gnns make effective predictions of signs in static signed networks~\cite{derr2018signed,huang2019signed,huang2021pole,LiuZ0ZCL021,ShuDC0ZXS21,HuangSHC21,LiTZC20}. 
}

\vspace{1mm}
\noindent\textbf{Challenges.} 
Modeling dynamic signed networks with existing (static) signed \gnns faces two challenges.

\textbf{(C1) Temporal-Awareness}, i.e., modeling the sequential order and evolution of signed interactions to learn node representations.
Static signed \gnns only learn the network structure at a given timestamp~\cite{derr2018signed,huang2019signed,huang2021pole}, ignoring the temporal order of the interactions. 
Additionally, while signed \gnns can incorporate balance theory in static networks, they are unable to follow balance theory when the network is dynamic~\cite{abell2009structural,hummon2003some,marvel2011continuous,doreian2019structural,wang2021inductive}. 

\textbf{(C2) Staleness}, i.e., in the absence of recent direct interactions of a node, its representation becomes stale. 
Staleness is a pervasive issue since real-world networks are long-tailed and sparse, i.e., most nodes only have a few direct interactions. 


Meanwhile, there has been a surge of interest in learning temporal representations of nodes in dynamic networks by developing dynamic GNN models~\cite{trivedi2019dyrep,kumar2019predicting,rossi2020temporal}. However, these model architectures are not \textbf{(C3) Sign-Aware}. 
In other words, since these models have been developed specifically for unsigned networks, they treat negative and positive edges the same and thus, fail to incorporate the higher-order organization induced by the link signs (i.e., structural balance theory). 
Due to their homophily assumptions in learning ~\cite{derr2018signed}, these models fail to model heterophilic network organizations such as negative links between polarized communities. 

\hide{Networks can evolve through different mechanisms. Both local and global changes can lead to making and breaking of links \srijan{why are you talking about this? your model does not distinguish between local vs global.}. Exogenous shocks can also cause a network to evolve in a certain manner \srijan{why are you talking about this? your model does nothing about exogenous shocks. Information for the sake of providing information is not needed.}. This is further complicated when links are signed since along with link prediction, sign prediction also becomes a problem. 
theory should be formulated in terms of a dynamical system since structural balance only describes an equilibrium state and does not provide any model on how the network evolves \cite{abell2009structural,hummon2003some,marvel2011continuous,doreian2019structural,wang2021inductive}. A learning model must therefore incorporate both balance and network evolution in order to model such networks. 
}


\vspace{1mm}
\noindent\textbf{Proposed Ideas.} In this work, we aim to \ul{effectively learn representations of dynamic signed networks while addressing temporal-awareness, staleness, and sign-awareness problems.}
Towards this goal, we propose a new dynamic signed GNN model, named as \textbf{\name}: \textbf{\ul{S}}igned link \textbf{\ul{E}}volution using \textbf{\ul{M}}emory modules and \textbf{\ul{B}}alanced \textbf{\ul{A}}ggregation. 
\name learns a short-term `memory' encoding and a long-term `embedding' per node. 
\name consists of three key modules: 
\begin{enumerate}[leftmargin=*]
    \item \textit{Signed Memories:} \name learns two signed memories (positive and negative) per node to distinctly encode past interactions of a node with its friends and enemies, respectively.
    \item \textit{Balanced Aggregation:} when the network evolves by adding a link ($u,v$), \name updates signed memories of nodes $u$ and $v$  while explicitly incorporating balance theory. 
    \item \textit{Long-term Propagation:} \name learns a node's embedding by attentively aggregating signed memories over all its neighbors until the current timestep. This overcomes staleness by learning not only from direct interactions at that time but any interactions in any of its past neighbors. 
\end{enumerate}

\hide{
In (M1), \name learns two signed memories (positive and negative) per node to distinctly encode past interactions of a node with its friends and enemies, respectively.
In (M2), when the network evolves by adding a link ($u,v$), \name updates signed memories of nodes $u$ and $v$  while explicitly incorporating balance theory. 
In (M3), \name builds a node's embedding by attentively aggregating signed memories of its $k$-hop neighborhood. This overcomes staleness by learning from all interactions within $k$-hop neighborhoods. 
}

Representations thus formed can be used to solve multiple different tasks on dynamic signed networks: predicting the existence of a future link (i.e., \textbf{dynamic link existence prediction}), predicting its sign (i.e., \textbf{dynamic link sign prediction}), predicting both existence and sign in future (i.e., \textbf{dynamic link signed existence prediction}), and predicting its signed weight (i.e., \textbf{dynamic link signed weight prediction}).
While dynamic link existence prediction has been studied extensively for unsigned dynamic networks in the literature~\cite{rossi2020temporal,wang2021inductive,chen2022gc}, no work has studied it for signed networks. Similarly, while sign and signed weight prediction have been studied for static signed networks~\cite{derr2018signed,huang2019signed,javari2020rose,huang2021pole,kumar2016edge}, no prior work exists that studies the dynamic setting.
Due to the inductive nature of the dynamic case, \name generalizes to new nodes and links, which has never been addressed in the signed network literature. 
We show experimentally that (1) modeling both balance and network evolution is essential to accurately predict signs and signed weights, and (2) \name consistently and significantly outperforms state-of-the-art baselines for signed networks and for dynamic networks. 
\noindent\textbf{Contributions.} Our contributions are summarized as follows: 
\begin{itemize}[leftmargin=*]
    \item \textbf{Novel Problem:} We devise a novel problem of representing dynamic signed networks via $4$ different prediction tasks.
    \item \textbf{Dynamic Signed \gnn:} We propose \name, a GNN-based model to represent dynamic signed networks.
    To the best of our knowledge, we are the first to devise a model to jointly address temporal-awareness, staleness, and sign-awareness problems.
    
    \item \textbf{Comprehensive Validation:} We validate the effectiveness of \name by comparing it with $5$ baselines on $4$ real-world datasets under the problems of dynamic link existence, sign, and signed weight prediction. Specifically, \name matches or outperforms the baselines in all tasks across datasets with an improvement of up to $80\%$ in predicting signs. 
    

\end{itemize}

\hide{It combines balance theory with an efficient memory module to represent evolving signed links in a graph. 
Each node accumulates its past positive and negative interactions in two separate signed memory modules, accumulating the positive and negative historical events at the node. Upon addition of a new link, these memories are updated using signed messages generated at each node. These messages are generated by aggregating the memories from both nodes depending on the sign of the interaction. The aggregation follows the rules of balance theory. 
Finally, graph attention modules are used to generate evolving embeddings from the concatenation of the two memories. 
These components are trained in an end-to-end manner to learn effective representations.}

\vspace{1mm}

\section{Related Work}\label{sec:related_work}
In this section, we briefly review existing works for static signed GNNs and dynamic unsigned GNNs. Table \ref{tab:salesmat} compares our model with the existing works. Unlike other models, \name uses an end-to-end \gnn-based model to generate node representations of dynamic signed networks. 
\begin{table}[tb]
    \centering
    \caption{Comparison of existing models with the proposed approach \name based on the challenges that they address. 
    }
    \label{tab:salesmat}
    \scalebox{0.92}{
    \begin{tabular}{c@{\hskip 0mm}c@{\hskip 3mm}c@{\hskip 3mm}c@{\hskip 3mm}c}
    \toprule
        Method & Temporal (C1) & Staleness (C2) & Signed (C3) \\ 
        \midrule
        \gcn \cite{kipf2016semi} & & & \\ 
        \sgcn \cite{derr2018signed}  & & & \cmark \\ 
        \sigat~\cite{huang2019signed} & & & \cmark \\ 
        \tgn \cite{rossi2020temporal} & \cmark & \cmark &   \\ 
        \rwlstm \cite{dang2018link} & \cmark & & \cmark \\ 
        \name (proposed) & \cmark & \cmark & \cmark 
        \\\bottomrule
    \end{tabular}
    }
     \vspace{-0.3cm}
\end{table}

\subsection{Static Signed \gnn}

\begin{figure}[tb]
    \centering
    \includegraphics[width=\linewidth]{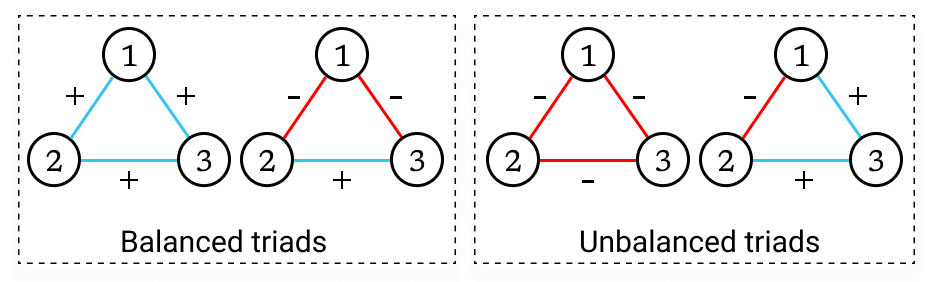}
    \vspace{-0.4cm}
    \caption{Balanced and unbalanced triads as dictated by balance theory~\cite{cartwright1956structural}.}
    \label{fig:balance}
    \vspace{-0.4cm}
\end{figure}


Machine learning on graphs has seen widespread interest, especially since the advent of message-passing \gnns.
Plenty of \gnns have been adapted to perform a variety of graph-related tasks~\cite{wu2020comprehensive}. 
Signed networks allow us to model polarity of the relationship which becomes particularly important in a variety of applications \cite{kumar2016edge,kumar2018rev2,west2014wikirfa}. Tasks of predicting the sign and signed weights of the links are often studied in the literature~\cite{derr2018signed,kumar2016edge}.
With the advent of \gnns, several works have developed \gnns for signed networks \cite{derr2018signed,huang2019signed,javari2020rose,huang2021pole}. They make use of the inductive bias induced by the neighborhood message passing mechanism, but augments it with ideas from social science like balance (as shown in Figure \ref{fig:balance}) \cite{cartwright1956structural} and status \cite{leskovec2010signed}.
However, most of these signed \gnns ignore the temporal aspect of the signed network and are limited to static networks.
As mentioned in Section~\ref{sec:intro}, this leads to temporal-awareness and staleness problems.
Since majority of networks in the real world are dynamic and the temporal interactions are rich sources of data, we present a model that exploits network evolution along with the signed information to make prediction on future signed links. 

\subsection{Dynamic Unsigned \gnn}

\hide{Dynamic graphs can be seen as a sequence of discrete snapshots over time as each timestamped event that happen between consecutive snapshots can be aggregated. Machine learning models have been developed to represent these structures by making use of the progress in sequential modeling through models like RNN-LSTM \cite{pareja2020evolvegcn}, attention \cite{sankar2020dysat}, etc.
}
Dynamic networks are popularly modeled as a stream of continuously time-stamped links,
and
allow us to capture the network evolution in real time. Methods like DyRep \cite{trivedi2019dyrep}, JODIE \cite{kumar2019predicting} and \tgn \cite{rossi2020temporal} have successfully modeled dynamic networks in continuous time, enabling them to solve multiple network-based problems. In particular, the problem of predicting whether a link would arrive in future has been studied extensively in the literature. GNNs are effective in this task as it is often the case that link dynamics is affected by the past interactions in the local neighborhood of the involved nodes. 

However, these models do not use the signed information, which we pointed out as the sign-awareness problem in Section~\ref{sec:intro}. 
In particular, signed interactions are shown to reorganize in well-defined higher-order structures such as balanced triads.
However, the existing works based on standard message-passing assume homophily and fail to capture the heterophilic nature of signed networks (e.g., polarized social networks). In this work, we incorporate balance theory to effectively learn representations of continuous-time dynamic networks. To the best of our knowledge, there has been only one prior work~\cite{dang2018link} in this direction, but not using \gnns. It uses random walks to model the network as an evolving time-series which is then fed into a recurrent neural network for sign prediction. However, it does not make use of balance theory, \gnns for processing the graph structure, or attention mechanism for processing temporal data. We shall build on these recent advancements to create a more accurate model. 



\section{Problem Definition}\label{sec:problem}

{\small
\begin{table}[t]
    \centering
    \label{tab:nots}
    \caption{Notations used in this paper}
    \renewcommand{\arraystretch}{1.2}
    \begin{tabular}{c|l}
    \toprule
     \textbf{Notation}                                                           &  \textbf{Description} \\ \midrule
        $\GG(t)$ & graph snapshot at $t$ \\
        $\GV(t)$ & vertex set at $t$ \\
        $\GE^+(t), \GE^-(t)$ & positive and negative edge set at $t$ \\
        $\GE(t)$ & $\GE^+(t) \cup \GE^-(t)$ \\
        $\GG_0$, $\GG_n$ & initial and final snapshot \\ \midrule
        $\ev$ & event like $\textsc{Add}_{edge}(u, v, \ebf_{uv})$ \\ 
        $\EE[t_i, t_f]$ & sequence of time-stamped events from $t_i$ till $t_f$ \\ 
        $\ebf_{uv}(t)$ & signed weight of $(u, v)$ at $t$ \\
        $\xbf_u(t)$ & node features of $u$ at $t$ \\
        \midrule
        $\Nbrs_u^k([0, t])$ & $k$-hop neighbor set of $u$ from time $0$ to $t$ \\
        $\sbf^+_u(t), \sbf^-_u(t)$ & positive/negative memory of node $u$ at $t$ \\
        $\mbf^+_u(t), \mbf^-_u(t)$ & positive/negative message encoding of $u$ at $t$ \\
        $\agmbf^+_u(t), \agmbf^-_u(t)$ & aggregated positive/negative messages on $u$ at $t$  \\ \midrule
        $B$ & Batch size  \\
        $T_B(t)$ & First time after $t$ with $B$ interactions in between \\
        $T_B^{-1}(t)$ & Last time from $t$ that was batched \\
        $p(u, v, t; \thetabf)$ & probability of edge $(u, v)$ for specified task at $t$ \\
        $\GE_{train}, \GE_{val}, \GE_{test}$ & training, validation, test set of edges \\
    \bottomrule 
    \end{tabular}
\end{table}
}

A directed continuous-time dynamic signed network 
at time $t$ is denoted by $\GG(t) = (\GV(t), \GE^+(t), \GE^-(t))$, where $\GV(t)$ denotes the set of nodes, and $\GE^+(t)$ (resp. $\GE^-(t)$) denotes the positive (resp. negative) links at time $t$. It
can be modeled using an initial network $\GG_0 = (\GV_0, \GE^+_0, \GE^-_0)$ along with a sequence of time-stamped events (ordered by time), $\EE := \EE[t_0, t_n] = \{(\tau, \ev): t_0 \le \tau \le t_n\}$ 
, such that $t_n$ is the final time step and $\ev$ represents a link addition at timestep $\tau$.\footnote{We only consider link additions in this work. Link removals and node-level events will be considered in future work.} More precisely, one can construct a network $\GG(t)$ at snapshot $t$ by augmenting links to $\GG_0$ such that $\GG(t_n) = \GG_0 \sqcup \{\ev \, : \; (t, \ev) \in \EE, t_0 \le t \le t_n\}$. The operator $\sqcup$ sequentially updates the network based on the events $\{\ev\}$. For example, if the event $\ev$ is addition of the link $(u, v, \ebf_{uv})$, i.e., $\textsc{Add}_{edge}(u, v, \ebf_{uv})$, and the link signed weight $\ebf_{uv}(t) > 0$, then network $\GG = (\GV, \GEp, \GEn)$ $\rightarrow (\GV, \GEp \cup \{(u, v, \ebf_{uv}(t))\}, \GEn)$. 
For the sake of simplicity, we will use $\GE(t) := \GEp(t) \cup \GEn(t)$ and $\GG_n$ to denote $\GG(t_n)$.



In this work, we study the effectiveness of signed links' representation in time through the following four problems:


\begin{prob}[\textbf{Dynamic Link Existence Prediction}]
    \label{prob:linkexpred}
    Given $\GG_0$, $\EE[t_1, t_j]$, the goal is to predict whether a link exists between $u$ and $v$ in future, i.e., $(u, v, \cdot)\in \GE(t_k)$ for $t_k$ > $t_j$.
\end{prob}

\begin{prob}[\textbf{Dynamic Link Sign Prediction}]
    \label{prob:signpred}
    Given $\GG_0$, $\EE[t_1, t_j]$,
    and a link $(u, v, \ebf_{uv}) \in \GE(t_k)$ for $t_k$ > $t_j$, the goal is to predict its sign, i.e., whether $\ebf_{uv}(t_k) > 0$ or $\ebf_{uv}(t_k) < 0$. 
\end{prob}

\begin{prob}[\textbf{Dynamic Link Signed Existence Prediction}]
    \label{prob:slinkpred}
    Given $\GG_0$, $\EE[t_1, t_j]$, the goal is to predict whether a positive, negative or no link exists between $u$ and $v$ in future, i.e., whether $(u, v, \cdot) \in \GE^+(t_k), (u, v, \cdot) \in \GE^-(t_k), $ or $(u, v, \cdot) \notin \GE(t_k)$ for $t_k$ > $t_j$.
\end{prob}

\begin{prob}[\textbf{Dynamic Link Signed Weight Prediction}]
    \label{prob:signwtpred}
    Given $\GG_0$, $\EE[t_1, t_j]$, 
    and a link $(u, v, \ebf_{uv}) \in \GE(t_k)$ for $t_k$ > $t_j$, the goal is to predict the edge weight $\ebf_{uv}(t_k)$.
\end{prob}

\section{\name: Our Approach}

In this section, we propose \name for representation learning in dynamic signed networks, which stands for \fname. 
We learn two representations per node: node memories to encode temporal signed interactions of the node, and 
node embeddings to incorporate interactions in its long-term neighbors. 
Finally, we use the learned embeddings to make the prediction. 

In Section~\ref{sec:overview}, we present an overview of \name. In Section~\ref{sec:pip}, we describe the four steps of \name in detail. In Section~\ref{sec:learning}, we discuss the process of learning and inference for \name. Lastly, in Section~\ref{sec:discussion}, we clarify our contributions in comparison to existing static signed and dynamic unsigned GNNs.

\subsection{Overview}\label{sec:overview}

\name is motivated by the following three observations that help to understand signed links evolution in a dynamic signed network:

\begin{enumerate}[leftmargin=*]
    \item \textbf{Links Evolution}: In a dynamic network, new links of a node $u$ at time $t$ tend to appear
    depending on the past interactions in the ego-network of $u$ \cite{trivedi2019dyrep,rossi2020temporal}, which consists of the focal node $u$ and its neighborhood to whom $u$ had a connection at some path length before $t$.
    \item \textbf{Structural Balance}: Balance theory~\cite{cartwright1956structural} guides signed network evolution.
    If $(u, v) \notin \GE$ but $(u, w), (w, v) \in \GE$, then the sign of $(u, w)$ and $(w, v)$ would determine how information from $v$ reaches $u$ and consequently, the sign of future link $(u, v)$. As a result, $u$ should consider the information coming from a friend of a friend (resp. a friend of an enemy) differently than the information coming from an enemy of a friend (resp. an enemy of an enemy).
    \item \textbf{Long-tailed Distribution}: It is well known that real-world networks tend to have a power-law degree distribution, which indicates only a few nodes in the network have a high degree, while most of the other nodes have a very low degree. In a dynamic network, it makes the representation of long-tailed nodes that have no \emph{recent} interactions more challenging. In other words, their representations can easily become stale.
    
    
\end{enumerate}


Following these observations, our model has three key architectural components --- signed memories, balanced aggregation, and long-term propagation. 

\textbf{Signed Memories.} A node's memory serves the purpose of representing the node's recent interactions with its immediate positive and negative neighbors, which is useful for predicting new links according to Observation 1. 
To be able to distinguish between positive versus negative neighbors as per Observation 2, our architecture uses two different memory modules, one positive and one negative, to accumulate the past interactions according to their sign. 

\textbf{Balanced Aggregation.}  Upon addition of a signed link $(u,v)$, both nodes' positive and negative memories must be updated based on Observation 2:

\begin{enumerate}[leftmargin=*]
    \item If link $(u,v)$ is positive,  memories of the \textit{same polarity} influence each other. Thus, the positive memory of $u$ is updated using the positive memory of $v$, while $u$'s negative memory is updated using the negative memory of $v$. 
    \item If link $(u,v)$ is negative, memories of \textit{opposite polarity} provide influence. Thus, the positive memory of $u$ is updated using the negative memory of $v$, while $u$'s negative memory is updated using the positive memory of $v$. 
\end{enumerate}

\begin{figure}[t]
    \centering
    \includegraphics[scale=1.2]{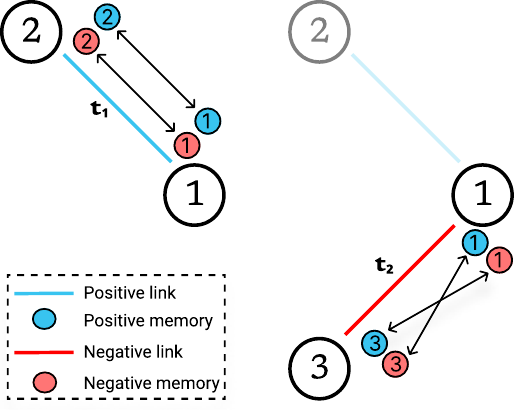}
     \vspace{-0.2cm}
    \caption{Balanced aggregation: Schema representing how link addition events \boldsymbol{$(t_1, (1, 2, +))$} and \boldsymbol{$(t_2, (1, 3, -))$} 
    update the corresponding signed memories at the end-nodes.
    }
     \vspace{-0.2cm}
    \label{fig:agg}
\end{figure}

Figure \ref{fig:agg} visualizes the balanced aggregation mechanism to update a node's memory due to incident positive and negative links. 

\begin{figure}[t]
    \centering
    \includegraphics[scale=1]{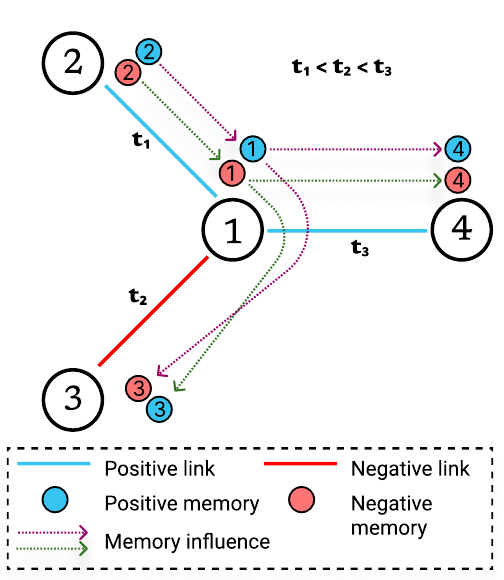} 
     \vspace{-0.2cm}
    \caption{Schema showing how updating memories based on balanced aggregation enforces balance theory in triadic relations \boldsymbol{$2-1-3$} and \boldsymbol{$2-1-4$} when \boldsymbol{$t_2 < t_1$}, \boldsymbol{$t_2 < t_3$}.
    }
    \label{fig:mem_dep}
    \vspace{-0.2cm}
\end{figure}

Furthermore, it should be noted that \textit{balanced aggregation naturally follows higher-order structural rules of balance theory}.
Figure \ref{fig:mem_dep} shows how updating node memories using neighbor-to-neighbor balanced aggregation results in complying with higher-order balance theory relationships. 
At the first time step $t_1$, node 2's positive memory updates node 1's positive memory, which then updates node 3's negative memory at the next time step $t_2$. 
Hence, one can note that positive memory from a friend of an enemy (i.e., node 2 is a friend of an enemy of node 3) updates the negative memory, which stores the information from enemies, thus, in line with balance theory. 
Similarly, other rules of balance theory can also be realized by local updates in future time steps. 
For example, node 2's positive memory updates node 4's positive memory via node 1's positive memory by the `friend of a friend is a friend' rule. 

\textbf{Long-Term Propagation.} A node's embedding serves the purpose of propagating signed memories of a node to the neighbors with no active interaction, which is useful for addressing staleness for long-tailed nodes according to Observation 3. 
To generate a node's embedding, we attentively aggregate signed memories of a node over all of its neighbors in the past and not just the neighbors in the current batch.




\subsection{\textbf{\name Model Pipeline}}\label{sec:pip}
\begin{figure*}[t]
    \centering
    \includegraphics[scale=0.6] 
    {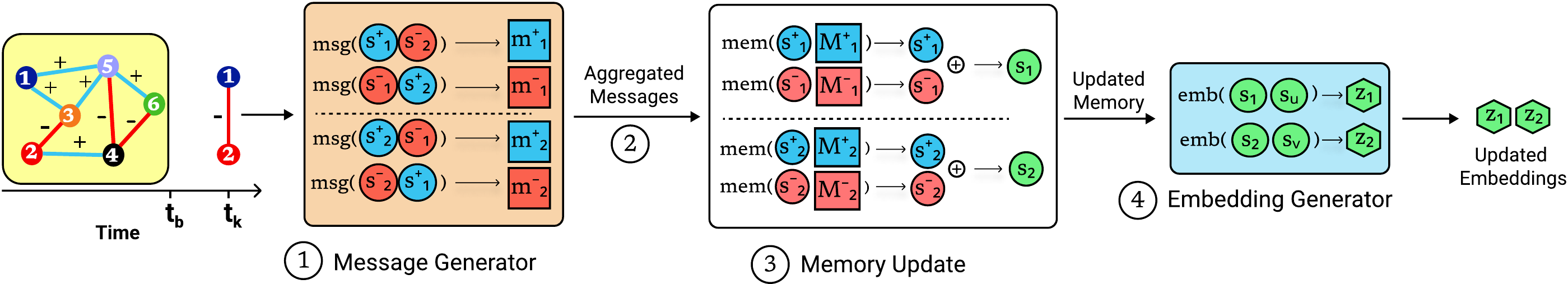}
    \caption{Pipeline of our method, \name. The encircled numbers dictate the sequence in which each step happens. 
    The model first
    adds the new link to the network, followed by generating messages as dictated by balance theory, in step 1. It then aggregates the messages and updates its memory using these new messages in steps 2 and 3. A graph attention based neural network is then used to update the embeddings of the nodes involved in step 4, which are used for downstream prediction tasks. 
    }
    \label{fig:model}
\end{figure*}

Figure \ref{fig:model} demonstrates the pipeline of our model. It consists of the following four steps: 
\begin{enumerate}[leftmargin=*]
    \item Encode the events as signed messages following balance theory
    \item Aggregate signed messages in a batch (for efficient processing)
    \item Update node memories with the aggregated messages
    \item Generate node embeddings using memories and other features
\end{enumerate}

\noindent\textbf{Temporal Batching.} Since networks evolve rapidly, it is computationally inefficient to update node embeddings after each change in the network. 
Thus, we combine a fixed number $B$ of sequential events happening after a time $t$  into a temporal batch. 
Node embeddings are then updated based on all the events included in such a batch. 
This also allows us to parallelize the computation of events within the batches and make the training process efficient. 
We create uniform batches of size $B$ starting from $t_0$. The first batch stores these $B$ interactions $\EE[t_0, T_B(t_0)]$ such that $T_B(x)$ is the smallest timestep after $x$ such that there are exactly $B$ events in $\EE[x, T_B(x)]$. Consequently, the $k$th batch consists of the events $\EE\left[T_B^{(k-1)}(t_0), T_B^{(k)}(t_0)\right]$ 
with $T_B^{(j)}(x) = T_B \circ T_B^{(j-1)}(x)$.


Now, we describe the details of each step. To this end, we consider the updates due to an event of the form $\ev = (t, \textsc{Add}(u, v, \ebf_{uv}))$. Let us also consider $T_B^{-1}(t)$ to be the latest batch time until $t$, i.e., $T_B^{-1}(t) = \max_k T_B^{(k)}(t_0)$ such that $T_B^{(k)} \le t$.
\subsection*{Step 1: Message Generation}
Given a new interaction between $u$ and $v$ by $\ev = (t, \textsc{Add}(u, v, \ebf_{uv}))$, we generate signed messages from $v$ to $u$, which contain $v$'s information passed to $u$, according to balance theory. The signed messages from $u$ to $v$ are also generated similarly.
Thus, we here explain how to generate the signed messages from $v$ to $u$. 

Specifically, if $e_{uv}$ is positive (resp. negative), then the message from positive (resp. negative) memories of node $v$ will be used by node $u$ to update $u$'s positive (resp. negative) memory. 
More formally, positive and negative messages from $v$ to $u$ at time $t$, $\mbf^+_{u\leftarrow v}(t)$ and $\mbf^-_{u\leftarrow v}(t)$, are generated as follows\footnote{$\mbf^+_{v\leftarrow u} (t)$ and $\mbf^-_{v\leftarrow u} (t)$ are also formed similarly.}:
\begin{equation}\label{eq:msgplus}
\begin{split}
    &\mbf^+_{u\leftarrow v} (t) = \\
    &\begin{cases}
        \msg(\sbf^+_u(t - \Delta t_u) \oplus \sbf^+_v(t - \Delta t_v) \oplus \Delta t_u \oplus |\ebf_{uv}(t)|), \ebf_{uv}(t) > 0,\\
        \msg(\sbf^+_u(t - \Delta t_u) \oplus \sbf^-_v(t - \Delta t_v) \oplus \Delta t_u \oplus |\ebf_{uv}(t)|), \ebf_{uv}(t) < 0,\\
    \end{cases}
\end{split}
\end{equation}
\begin{equation}\label{eq:msgminus}
\begin{aligned}
    &\mbf^-_{u\leftarrow v} (t) = \\
    &\begin{cases}
        \msg(\sbf^-_u(t - \Delta t_u) \oplus \sbf^-_v(t - \Delta t_v) \oplus \Delta t_u \oplus |\ebf_{uv}(t)|), \ebf_{uv}(t) > 0,\\
        \msg(\sbf^-_u(t - \Delta t_u) \oplus \sbf^+_v(t - \Delta t_v) \oplus \Delta t_u \oplus |\ebf_{uv}(t)|), \ebf_{uv}(t) < 0,\\
    \end{cases}
\end{aligned}
\end{equation}
\normalsize
where $\oplus$ indicates the concatenation operator and $\sbf^+_u$ (resp. $ \sbf^-_u$) denotes positive (resp. negative) memory of $u$. 
Also, $t - \Delta t_u$ denotes the time when the signed memories of node $u$ were last updated (i.e., the last time when there was a direct interaction involving the node $u$). 
In addition, $|\ebf_{uv}(t)|$ indicates the link signed weight between $u$ and $v$.
Lastly, we employ a standard multi-layer perception (MLP) to implement the message function \msg.

Intuitively, $\mbf^+_{u\leftarrow v} (t)$ and $\mbf^-_{u\leftarrow v} (t)$ preserve the information received through recent positive and negative memories of both $u$ and $v$ while taking into consideration of the balance theory. Furthermore, it considers the strength of the interaction between $u$ and $v$. 
By doing so, \name can (1) differentiate the impact of recent versus older interactions, (2) treat signed memories differently based on the link sign, and (3) differentiate the influence per link.
That is, signed messages are designed so that \name can jointly address the temporal-awareness (C1) and sign-awareness (C3) problems.

\subsection*{Step 2: Message Aggregation}
Given a set of positive and negative messages, i.e., $\mbf^+_{u\leftarrow (\cdot)} (t)$ and $\mbf^-_{u\leftarrow (\cdot)} (t)$, received by each node $u$ in the batch, we aggregate them separately as follows:
\begin{equation}\label{eq:msgagg}
\begin{aligned}
    \agmbf^+_u(t) &= \agg (\{\mbf^+_{u\leftarrow i}: \;(\tau, \textsc{Add}(i,u,\ebf_{iu})) \in \EE[T_B^{-1}(t), t]\}), \\
    \agmbf^-_u(t) &= \agg (\{\mbf^-_{u\leftarrow i}: \;(\tau, \textsc{Add}(i,u,\ebf_{iu})) \in \EE[T_B^{-1}(t), t]\}),
\end{aligned}
\end{equation}
To implement the aggregation function \agg, we can employ any methods used in \cite{rossi2020temporal}, such as mean aggregation, temporal aggregation using long short-term memories (LSTMs), and a function that returns the most recent message from its input set of messages.
We found no significant difference among these and used the function returning the most recent message.


\subsection*{Step 3: Memory Update}
Now, we have signed messages, i.e., $\agmbf^+_u(t)$ and $\agmbf^-_u(t)$, which encode new knowledge from the events involving each node $u$ in the batch from $\EE[T_B^{-1}(t), t]$
Given $\agmbf^+_u(t)$ and $\agmbf^-_u(t)$, we update $u$'s positive and negative memories, $\sbf^+_u (t)$ and $\sbf^-_u (t)$, at time $t$ and then form a joint memory $\sbf_u(t)$ for the node $u$ as follows: 
\begin{align}\label{eq:memupd}
    \sbf^+_u(t) &= \mem (\agmbf^+_u(t), \sbf^+_u(t - \Delta t)), \\ \nonumber
    \sbf^-_u(t) &= \mem (\agmbf^-_u(t), \sbf^-_u(t - \Delta t)), \\ \nonumber
    \sbf_u(t) &= \sbf^+ (t) \oplus \sbf^- (t),
\end{align}
where \mem indicates a memory update function, which we employ as an LSTM in this paper.


\subsection*{Step 4: Embedding Generation}

Given a joint memory $\sbf_u(t)$ of $u$ and memories $\sbf_i(t)$ of all of its neighbors $i$ until that time $t$, i.e., $i \in \Nbrs_u([0, t])$, we find each node $u$'s embedding $\zbf_u(t)$ at time $t$ using Graph Transformer~\cite{shi2020masked} as:
\begin{equation}\label{eq:emb}
    \zbf_u(t) = \Wbf_1\sbf_u(t) + \sum_{\mathclap{\substack{(\tau, \ev) \in \EE[0, t]\\ \ev = \textsc{Add}(i, u, \ebf_{iu})}}}{\alpha_{i,u}(\Wbf_2\sbf_i(t) + \Wbf_4 \tau + \Wbf_4|\ebf_{iu}|)},
\end{equation}
where the attention coefficients $\alpha_{i,u}$ are found using multi-head dot product attention, and $\Wbf_j$ denote learnable weights.
If there are node features $\xbf_u(t)$, we concatenate them with the node memories $\sbf_u(t)$ and pass the concatenated value in place of $\sbf_u(t)$ in Eq.~(\ref{eq:emb}) to obtain the embeddings $\zbf_u(t)$.


Therefore, one can note that $u$'s embedding $\zbf_u$ will be updated at time $t$ even if there are no direct interactions with $u$ in the batch. This is due to the fact that we attentively aggregate the memories of all of its neighbors formed until now and not just in this batch. Thus, if any of these nodes have direct interactions in this batch, their memories would be updated, subsequently updating $u$'s embedding. In this way, \name can generate robust node embeddings even in the lack of direct interactions during a time period, thereby successfully mitigating the staleness problem (C2).

\hide{To make effective predictions about events in a batch, the node representations are first updated using events until the previous temporal batch. 
The predictions are made using these latest embeddings, which ensures that the representations accounts for the immediate history. 
These representations are then used for updates when sequentially processing the events within the batch.
}

\subsection{Learning and Inference}\label{sec:learning}
To make effective predictions about events, the node embeddings are first updated using events until the previous temporal batch. 
The predictions are made using these latest embeddings, which ensures that the embeddings account for the immediate history. 
The node embeddings are used to make inference on a node pair $(u, v)$ at a time $t$ by concatenating the corresponding embeddings $\zbf_u(t)$ and $\zbf_v(t)$ and then passing it through an \mlp, i.e.,
\begin{equation}
    \widehat{y}(u, v, t; \thetabf) = \mlp (\zbf_u(t) \oplus \zbf_v(t)), 
\end{equation}
where $\thetabf$ denotes the set of learnable model parameters.

We treat Dynamic Link Existence and Dynamic Link Sign Prediction (Problems~\ref{prob:linkexpred},~\ref{prob:signpred}) as binary classification tasks. For Problem~\ref{prob:linkexpred}, negative sampling was employed to find an equal number of non-existing edges. To train the models for these tasks, we minimize the Binary Cross Entropy (BCE) loss in sequential batches as:
\begin{equation}\label{eq:loss_bce}
    \Ls_{BCE} (\GE_{tr}; \thetabf) = -\sum_{\mathclap{(u, v, t) \in \GE_{tr}}} \log p_y (u, v, T_B^{-1}(t); \thetabf)),
\end{equation}
where $\GE_{tr}$ denotes the training set and $p_y := \sigma(\widehat{y})$ if $y_{uv}(t) = 1$ and $1 -  \sigma(\widehat{y})$ otherwise. Note that $T_B^{-1}(t)$ is the latest batch time until $t$ and $y_{uv}(t)$ is the actual label of the edge, which is $1$ for link existence in Problem~\ref{prob:linkexpred} and $1$ for positive link in Problem~\ref{prob:signpred}.

Dynamic Signed Link Prediction (Problem~\ref{prob:slinkpred}) is treated as a multi-class classification task. Here, we minimize the Cross-Entropy Loss similar to Equation~\ref{eq:loss_bce} with the true labels are in $\{0, 1, 2\}$ denoting positive, negative, and non-existing edges respectively. Corresponding $p_y$ is found after applying softmax of $\widehat{y}$.

Finally, we consider the Dynamic Link Signed Weight Prediction (Problem~\ref{prob:signwtpred}) as a regression task for which we minimize the Root Mean Squared Error (RMSE) in sequential batches as follows:
\begin{equation}\label{eq:loss_rmse}
    \Ls_{RMSE} (\GE_{tr}; \thetabf) = \sqrt{\text{\footnotesize $\frac{1}{|\GE_{tr}|}$}\sum_{\hspace{5mm}\mathclap{(u, v, t) \in \GE_{tr}}} (\ebf_{uv}(t) - \widehat{y}(u, v, T_B^{-1}(t); \thetabf))^2}.
\end{equation}

\subsection{Discussion}\label{sec:discussion}

In this subsection, we summarize the unique architectural contribution of \name as compared to closely related \tgn~\cite{rossi2020temporal}. \tgn is a temporal graph learning module specifically designed for homophilic unsigned networks. On the other hand, the use of two different memories and their update rules inspired by balance theory, \name makes the learning appropriate to heterophilic signed networks. As we shall see in the experiments, such simple modification can be very effective in learning representations of links in a continuous-time dynamic signed network, by jointly addressing the temporal awareness, staleness, and sign-awareness problems.

\section{Experiments}
We designed our experiments, aiming at answering the following key evaluation questions (EQs):
\begin{itemize}[leftmargin=*]
    \item \textbf{EQ1}: How does \name perform compared to its competitors on the sign and link prediction tasks (Problems~\ref{prob:linkexpred},~\ref{prob:signpred},~\ref{prob:signwtpred} and ~\ref{prob:slinkpred})?
    \item \textbf{EQ2}: How does the performance vary in transductive (nodes seen during training) and inductive (unseen nodes) settings? 
    \item \textbf{EQ3}: How important is each component of \name?
    \item \textbf{EQ4}: Is \name time and memory efficient?
\end{itemize}

\subsection{Experimental Setup}

\begin{table}[tb]
    \centering
    \caption{Summary of the datasets. $f_+ = |\GE^+_f|/|\GE_f|$ denotes the fraction of positive links and $f_{ub} = \bigtriangleup_{ub}(\GG_f)/\bigtriangleup(\GG_f)$ denotes the fraction of unbalanced triangles in the signed network. Weight denotes the presence of link weight in a dataset.}
    \vspace{-0.2cm}
    \begin{tabular}{ l c @{\hskip 2mm} c @{\hskip 2mm} c @{\hskip 2mm} c @{\hskip 2mm} c @{\hskip 2mm} c}
    \toprule
    \textbf{Dataset} & \textbf{\#nodes} & \textbf{\#links} & \boldsymbol{$f_{+}$} & \boldsymbol{$f_{ub}$} &
    \textbf{\#days} & \textbf{Weight} \\
    \midrule
    
    \btcotc & 5.9K & 21.5K & 0.89 & 0.13 & 1,905 & \cmark \\ 
    \btcalpha & 3.8K & 14.1K & 0.85 & 0.14 & 1,902 & \cmark \\ 
    \wikirfa & 12.8K & 170.5K & 0.77 & 0.24 & 3,758 & \\ 
    \epinions & 131.9K & 871.4K & 0.85 & 0.09 & 945 & \\ 
    \bottomrule
    \end{tabular}
    \vspace{-0.2cm}
    \label{tab:dataset_stats}
\end{table}

\subsubsection{Datasets}
We used four publicly available\footnote{\url{https://snap.stanford.edu/data/index.html\#signnets}} real-world signed networks with temporal links: \btcotc, \btcalpha, \wikirfa, and \epinions. Table~\ref{tab:dataset_stats} shows some statistics of these datasets.  
\begin{itemize}[leftmargin=*]
    \item \textbf{\btcotc} and \textbf{\btcalpha}~\cite{kumar2018rev2,kumar2016edge} are two who-trusts-whom networks of people trading Bitcoin on OTC and Alpha platforms, respectively. In addition to the sign and timestamp, these links are also weighted on a scale from $-10$ to $+10$. 
    \item \textbf{\wikirfa} \cite{west2014wikirfa} is a voting network by fellow members for a Wikipedia member's request of adminship. 
    \item \textbf{\epinions} is a trust network between members of an online consumer review site.
\end{itemize}

From Table~\ref{tab:dataset_stats}, we note that all datasets are skewed towards positive links (see $f_{+}$) and follow balance theory (see $f_{ub}$). 
Note that \wikirfa originally has links with null signs that denote abstention, 
which we filtered out 
to make it into a binary signed network. 
Furthermore, we also removed any links with no given timestamps or no given weights from all the datasets.
Since no datasets contain node features, we use a zeroed vector of dimension $8$ as the features.




\subsubsection{Baselines}
We study a novel problem on dynamic signed networks that have been studied in only one prior work, i.e., \rwlstm~\cite{dang2018link}; we made its variant  \textbf{\sgclstm}, which can be used in the inductive setting for a fair compariosn.
Also, we employ four state-of-the-art GNNs as additional baselines:
one static unsigned GNN, i.e., \textbf{\gcn}~\cite{kipf2016semi}; two static signed GNN, i.e., \textbf{\sgcn}~\cite{derr2018signed} and \textbf{\sigat}~\cite{huang2019signed}; and one dynamic unsigned GNN, i.e., \textbf{\tgn}~\cite{rossi2020temporal}. 
\begin{itemize}[leftmargin=*]
    \item \textbf{\gcn}~\cite{kipf2016semi} uses neighborhood convolution to generate node embeddings for static unsigned networks.
    \item \textbf{\sgcn}~\cite{derr2018signed} incorporates the balance theory in the neighborhood aggregation for static signed networks.
    \item \textbf{\sigat}~\cite{huang2019signed} generalizes GATs to incorporate motifs guided by balance and status theory. 
    \item \textbf{\sgclstm}. Since the existing work \rwlstm~\cite{dang2018link} does not use a GNN, it cannot be used in our inductive setting where unseen nodes may appear during the test phase. For a fair comparison, we thus employ an inductive \sgcn module in place of the random walks. Thus, \sgclstm combines \sgcn and \lstm modules sequentially such that graphs in each batch are encoded using \sgcn and these embeddings are passed as a sequence to an \lstm to obtain the representations. Thus,  \sgclstm would act like a discrete-time \textit{time-and-graph} model~\cite{gao2021equivalence}.
    \item \textbf{\tgn}~\cite{rossi2020temporal} is the state-of-the-art model to make predictions on dynamic unsigned networks. It encodes unsigned temporal interactions in a memory encoding and uses these to effectively learn temporal representations of nodes. 
\end{itemize}





\subsubsection{Evaluation Metrics} We use appropriate metrics to evaluate the performance on the $4$ different tasks, specified in Section~\ref{sec:problem}.

 For the binary classification tasks (\textit{Dynamic Link Sign Prediction} and \textit{Dynamic Link Existence Prediction}), we use the \fone and \auc scores between the true and predicted labels/probabilities. \fone measures the harmonic mean of the precision and recall of the positive class, while \auc computes the Area Under the Receiver Operating Characteristic (Precision vs Recall) curve drawn at different thresholds. While \fone handles the imbalance of classes in the data, \auc ignores the class imbalance. Higher \fone and \auc values are desired from a good classifier.

 For multiclass classification, i.e., the \textit{dynamic link signed existence Prediction}, we aggregate the \fone metric over different classes as an unweighted mean or macro averaging (\fonemac), an average weighted by the support (\fonewt), and a global calculation of precision/recall (\fonemic or equivalently, accuracy \acc). \fonewt accounts for imbalance in classes by weighing based on the number of samples while \fonemac gives equal weight to all the classes. On the other hand, \acc may perform well by just predicting the majority class.

 For the regression task, i.e., the \textit{Dynamic Link Signed Weight Prediction}, we use the \rmse, \rsq, and \kldiv between the predicted and actual signed weights. \rmse measures the root mean squared errors between the actual and predicted values while \rsq measures the proportion of variation in the outcome that is explained by the predicted value. Finally, \kldiv measures KL-divergence between the actual and predicted distributions. We expect a high \rsq and a low \rmse, \kldiv values from a good model.

\subsubsection{Implementation Details}
To evaluate prediction quality in the future, we order all the links according to their timestamps and then spilt them into training (70\%), validation (15\%), and test (15\%) sets.
All models are trained using a common framework described in Section~\ref{sec:learning} with the Adam optimizer~\cite{kingma2014adam}.
The predictions are made in a temporal batch, i.e., $t_k = T_B(t_j)$, where the batch size $B$ is fixed to be $1,000$ interactions for \bitcoinotc, \bitcoinalpha, and \wikirfa, and $16,000$ interactions for the larger \epinions dataset.
Note that the evaluation is also done in sequential batches, i.e., to predict a link at time $t$, all events (including those in the test set)
before $T_B^{-1}(t)$ are included in the model's context for prediction. 
In other words, while the parameters remain fixed during the test phase, embeddings are updated in an online manner to incorporate past events to predict the future. For static models, i.e., \gcn, \sgcn, and \sigat, we thus pass the graph formed by all the interactions until the current batch, i.e., $\GG(t)$ for prediction.

We carefully tune the hyperparameters such as the learning rate and the number of epochs of baselines for each method and dataset via grid search on the validation set. The embedding dimension is fixed to be $64$ and the transformer has $8$ attention heads.
All the experiments were conducted on the NVIDIA Tesla K$80$ GPUs with $12$GB memory, and the models were implemented using the PyTorch Geometric framework.\footnote{\url{https://github.com/pyg-team/pytorch_geometric}}  


\subsection{Prediction Performance}\label{sec:performance}
\begin{table}[tb]
    \centering
    \caption{Performance on dynamic link existence prediction (rounded to $2$ decimal digits). Higher \auc and \fone are desired. Bold values are the best performing and underlined values are the second best. `\oom' denotes out of GPU memory. 
    }
    \vspace{-0.2cm}
    \renewcommand{\arraystretch}{1}
    \resizebox{0.49\textwidth}{!} {
    \begin{tabular}{c | c c c c c c c c}
    \toprule
    \multirow{2}{*}{\textbf{Method}} & \multicolumn{2}{c}{\textbf{\btcotc}} & \multicolumn{2}{c}{\textbf{\btcalpha}} & \multicolumn{2}{c}{\textbf{\wikirfa}} & \multicolumn{2}{c}{\textbf{\epinions}} \\
    & \fone & \auc & \fone & \auc & \fone & \auc & \fone & \auc \\
    \midrule
    \gcn & 0.00 & 0.49 & 0.00 & 0.50 & 0.00 & 0.49 & 0.00 & 0.49 \\
    
    \sgcn & 0.75 & 0.82 & 0.79 & 0.83 & 0.64 & 0.75 & 0.81 & 0.88 \\

    \sigat & 0.77 & \ul{0.87} & \ul{0.85} & \ul{0.93} & \ul{0.86} & \ul{0.92} & \oom & \oom \\

    
    \sgclstm & 0.25 & 0.57 & 0.31 & 0.59 & 0.50 & 0.33 & \oom & \oom \\ 
    
    \tgn & \textbf{0.97} & \textbf{0.99} & \textbf{0.98} & \textbf{0.99} & \textbf{0.99} & \textbf{0.99} & \ul{0.94} & \ul{0.98} \\
    
    \midrule
    \textbf{\name} & \ul{0.96} & \textbf{0.99} & \textbf{0.98} & \textbf{0.99} & \textbf{0.99} & \textbf{0.99} & \textbf{0.95} & \textbf{0.99} \\
    \bottomrule
    \end{tabular}
    }
    \label{tab:linkexpred}
\end{table}

\begin{table}[tb]
    \centering
    \caption{Performance on dynamic link sign prediction (rounded to $2$ decimal digits). Higher \auc and \fone are desired. Bold values are the best performing and underlined values are the second best. `\oom' denotes out of GPU memory.
    }
    \vspace{-0.2cm}
    \renewcommand{\arraystretch}{1}
    \resizebox{0.49\textwidth}{!} {
    \begin{tabular}{c | c c c c c c c c}
    \toprule
    \multirow{2}{*}{\textbf{Method}} & \multicolumn{2}{c}{\textbf{\btcotc}} & \multicolumn{2}{c}{\textbf{\btcalpha}} & \multicolumn{2}{c}{\textbf{\wikirfa}} & \multicolumn{2}{c}{\textbf{\epinions}} \\
    & \fone & \auc & \fone & \auc & \fone & \auc & \fone & \auc \\
    \midrule
    \gcn & 0.03 & 0.54 & 0.06 & 0.51 & 0.00 & 0.52 & 0.00 & 0.47 \\
    
    \sgcn & 0.67 & 0.65 & 0.72 & 0.61 & 0.42 & 0.64 & 0.22 & 0.64 \\

    \sigat & 0.39 & 0.70 & 0.49 & \ul{0.71} & \ul{0.55} & \ul{0.68} & \oom & \oom \\

    
    \sgclstm & 0.12 & 0.54 & 0.12 & 0.56 & 0.00 & 0.05 & \oom & \oom \\ 
    
    \tgn & \ul{0.79} & \ul{0.79} & \ul{0.73} & 0.70 & 0.45 & 0.62 & \ul{0.38} & \ul{0.67} \\
    
    \midrule
    \textbf{\name} & \textbf{0.80} & \textbf{0.83} & \textbf{0.79} & \textbf{0.76} & \textbf{0.81} & \textbf{0.85} & \textbf{0.57} & \textbf{0.80} \\
    \bottomrule
    \end{tabular}
    }
    \label{tab:linksignpred}
\end{table}

To test the effectiveness of representations learned by \name, we evaluate its performance against representative baselines on the four tasks outlined earlier in Section~\ref{sec:problem}.

\begin{table*}[tb]
    \centering
    \caption{Performance on the $3$-class dynamic link signed existence prediction (rounded to $2$ digits). Higher \fonewt, \fonemac, and \acc are desired. Bold values are the best performing and underlined values are the second best. `\oom' denotes out of GPU memory.
    }
    \vspace{-0.2cm}
    \renewcommand{\arraystretch}{1}
    \begin{tabular}{c | c c c c c c c c c c c c}
    \toprule
    \multirow{2}{*}{\textbf{Method}} & \multicolumn{3}{c}{\textbf{\btcotc}} & \multicolumn{3}{c}{\textbf{\btcalpha}} & \multicolumn{3}{c}{\textbf{\wikirfa}} & \multicolumn{3}{c}{\textbf{\epinions}} \\
    & \fonewt & \fonemac & \acc & \fonewt & \fonemac & \acc & \fonewt & \fonemac & \acc & \fonewt & \fonemac & \acc \\
    \midrule
    \gcn & 0.37 & 0.30 & 0.34 & 0.37 & 0.29 & 0.33 & 0.39 & 0.31 & 0.37 & 0.35 & 0.30 & 0.33 \\
    
    \sgcn & 0.65 & 0.53 & \ul{0.63} & \ul{0.70} & 0.57 & 0.69 & 0.57 & 0.48 & 0.59 & 0.71 & 0.55 & 0.74  \\

    \sigat & \ul{0.67} & \ul{0.55} & 0.62 & \ul{0.70} & 0.57 & 0.65 & 0.65 & 0.55 & 0.61 & \oom & \oom & \oom  \\

    
    \sgclstm & 0.40 & 0.32 & 0.51 & 0.41 & 0.35 & 0.52 & 0.33 & 0.22 & 0.49 & \oom & \oom & \oom  \\
    
    \tgn & \textbf{0.88} & \textbf{0.76} & \textbf{0.87} & \textbf{0.86} & \ul{0.72} & \ul{0.84} & \ul{0.85} & \ul{0.71} & \ul{0.83} & \ul{0.79} & \ul{0.68} & \ul{0.76} \\
    
    \midrule
    \textbf{\name} & \textbf{0.88} & \textbf{0.76} & \textbf{0.87} & \textbf{0.86} & \textbf{0.73} & \textbf{0.85} & \textbf{0.91} & \textbf{0.80} & \textbf{0.90} & \textbf{0.82} & \textbf{0.74} & \textbf{0.81}  \\
    \bottomrule
    \end{tabular}
    \label{tab:slinkpred}
\end{table*}

\begin{table}[tb]
    \centering
    \caption{Performance (rounded to $2$ decimal digits) on dynamic link signed weight prediction. Lower \rmse and \kldiv and higher \rsq are desired. Bold values are the best performing and underlined values are the second best. 
    }
    \vspace{-0.2cm}
    \renewcommand{\arraystretch}{1}
    \resizebox{0.45\textwidth}{!} {
    \begin{tabular}{c | c c c c c c}
    \toprule
    \multirow{2}{*}{\textbf{Method}} & \multicolumn{3}{c}{\textbf{\btcotc}} & \multicolumn{3}{c}{\textbf{\btcalpha}} \\
    & \rmse & \rsq & \kldiv & \rmse & \rsq & \kldiv \\
    \midrule
    \gcn & 1.85 & -0.01 & 21.44 & 1.94 & -0.02 & 11.45  \\
    
    \sgcn & \ul{1.84} & \ul{-0.00} & 21.44 & \ul{1.93} & \ul{-0.00} & 20.82  \\

    \sigat & \ul{1.84} & \ul{-0.00} & 21.44 & 1.94 & -0.01 & 20.82  \\

    
    \sgclstm & \ul{1.84} & \ul{-0.00} & 21.44 & 1.94 & \ul{-0.00} & 20.82  \\
    
    \tgn & \textbf{1.83} & \textbf{0.01} & \ul{9.42} & \textbf{1.91} & \textbf{0.02} & \ul{9.34}  \\
    
    \midrule
    \textbf{\name} & \textbf{1.83} & \textbf{0.01} & \textbf{8.20} & \ul{1.92} & \textbf{0.02} & \textbf{8.93}  \\
    \bottomrule
    \end{tabular}
    }
    \label{tab:linkswtpred}
\end{table}

\subsubsection{Dynamic Link Existence Prediction.} Table~\ref{tab:linkexpred} shows the performance of different methods on the task of predicting whether a link exists between two nodes at a time step $t$ given the graph until the previous batch $T_B^{-1}(t)$. This is a standard task in dynamic graphs literature~\cite{rossi2020temporal,xu2020inductive}. We find that \name closely matches the performance of state-of-the-art \tgn in this task for signed networks with a maximum difference of $0.01$ in the \fone and \auc metrics. This shows that \name can effectively handle the evolution of link format in dynamic signed networks. \sigat is the second closest baseline with a difference of at least $0.13$ \fone and at least $0.06$ \auc points. Thus, we conclude that dynamic \gnns (such as \tgn and \name) are essential to model link existence in dynamic signed networks as they show an average of $14.2\%$ and a minimum of $6.4\%$ improvement over their static counterparts. We also find that \sgclstm is highly ineffective in modeling link formation dynamics and only shows a slight improvement over static unsigned \gcn model, which produces random results. We know from the literature that \textit{time-and-graph} models are theoretically less expressive than \textit{time-then-graph} models~\cite{gao2021equivalence}. Our results complement this theory by showing that even a simple \textit{time-then-graph} model that aggregates all the temporal interactions in a single graph and pass the aggregated graph to a static signed \gnn (as is done for \sgcn and \sigat baselines) is more effective than a \textit{time-and-graph} model such as \sgclstm. More sophisticated \textit{time-then-graph} models like \tgn and \name are found to be even more effective.

\subsubsection{Dynamic Link Sign Prediction} Table~\ref{tab:linksignpred} shows the performance of different methods on the task of predicting whether the sign of a future link at $t$ is positive or negative given the graph until $T_B^{-1}(t)$. We find that \name outperforms the baselines across all the datasets with an improvement of up to $80\%$ in the \fone measure and up to $33\%$ in the \auc measure. Note that due to the scarcity of negative signs in the datasets, there is a class imbalance (see Table~\ref{tab:dataset_stats}). Thus, \fone is a better measure to assess the quality of the classifier. \tgn is the second best performing method, closely followed by the static signed \sigat and \sgcn models. This shows that sign classification requires effective modeling both temporal (recall C1) and sign (C3) features, which is achieved jointly by \name. As shown by the suboptimality of \sigat and \tgn models through our results, neither balance theory nor temporal modeling alone can predict the signs of links in the future. Here too, we see poor performance of \sgclstm which can be attributed, as earlier, to its \textit{time-and-graph} modeling.

\subsubsection{Dynamic Link Signed Existence Prediction} Here, we combine the previous two tasks to do a $3$-class classification, i.e., to predict whether a link exists between two nodes in the future and if it exists, predict its sign. Table~\ref{tab:slinkpred} compares the performance of different methods for this task. \name outperforms or matches the baselines across all the datasets with an improvement of up to $7\%$ in the weighted \fone (\fonewt), $12.6\%$ in the macro \fone (\fonemac), and $8.4\%$ in the micro \fone (or accuracy \acc) scores. \tgn is the closest baseline and matches \name's performance on \btcotc and \btcalpha datasets. However, \name shows significant improvements over \tgn for the larger \wikirfa and \epinions datasets. Note that \fonemac takes a mean by giving equal weights to all three classes while \fonewt weighs each class by the number of samples it has. More improvement in \fonemac than \fonewt would then mean more improvement in the minority (i.e., negative signed links) class. Thus, we can note that \name specifically improves over \tgn by improving on classifying node pairs with a negative link, owing to the implicit prior of balance theory in its architecture. Here too, we note the poor performance of \sgclstm for similar reasons as mentioned above. Furthermore, we find that static signed methods (\sgcn and \sigat) are highly suboptimal for this task as the temporal methods (\tgn and \name) outperform them by a margin of at least $15.5\%$ in \fonewt, at least $28.1\%$ in \fonemac, and $9.5\%$ in \acc measures. This shows that modeling temporal evolution (recall C1) is essential for predicting signed structure of graphs in future.

\begin{table*}[tb]
    \centering
    \caption{Performance (rounded to $2$ decimal digits) on the $3$-class dynamic link signed existence prediction for transductive (no new nodes) and inductive (both new nodes) settings. Higher \fonewt and \fonemac is desired. Bold values are the best performing and underlined values are the second best. `\oom' denotes out of GPU memory.}
    \vspace{-0.2cm}
    \renewcommand{\arraystretch}{1}
    \resizebox{0.98\textwidth}{!} {
    \begin{tabular}{c | c c c c c c c c c c c c c c c c}
    \toprule
    \multirow{3}{*}{\textbf{Method}} & \multicolumn{4}{c}{\textbf{\btcotc}} & \multicolumn{4}{c}{\textbf{\btcalpha}} & \multicolumn{4}{c}{\textbf{\wikirfa}} & \multicolumn{4}{c}{\textbf{\epinions}} \\
    & \multicolumn{2}{c}{Transductive} & \multicolumn{2}{c}{Inductive} & \multicolumn{2}{c}{Transductive} & \multicolumn{2}{c}{Inductive} & \multicolumn{2}{c}{Transductive} & \multicolumn{2}{c}{Inductive} & \multicolumn{2}{c}{Transductive} & \multicolumn{2}{c}{Inductive}  \\
    & \fonewt & \fonemac & \fonewt & \fonemac & \fonewt & \fonemac & \fonewt & \fonemac & \fonewt & \fonemac & \fonewt & \fonemac & \fonewt & \fonemac & \fonewt & \fonemac \\
    \midrule
    \gcn & 0.38 & 0.31 & 0.36 & 0.28 & 0.34 & 0.28 & 0.37 & 0.29 & 0.33 & 0.29 & 0.42 & 0.34 & 0.35 & 0.30 & 0.36 & 0.29 \\
    
    \sgcn & 0.69 & 0.59 & 0.77 & 0.65 & \ul{0.75} & \ul{0.69} & 0.79 & \ul{0.67} & 0.60 & 0.50 & 0.67 & 0.57 & 0.75 & 0.57 & \textbf{0.84} & \textbf{0.76} \\

    \sigat & 0.68 & 0.58 & 0.76 & 0.63 & 0.70 & 0.65 & 0.74 & 0.61 & 0.60 & 0.52 & \ul{0.71} & \ul{0.62} & \oom & \oom & \oom & \oom \\

    
    \sgclstm & 0.46 & 0.39 & 0.32 & 0.21 & 0.50 & 0.44 & 0.33 & 0.26 & 0.33 & 0.22 & 0.33 & 0.22 & \oom & \oom & \oom & \oom \\
    
    \tgn & \ul{0.82} & \ul{0.68} & \ul{0.84} & \ul{0.68} & \ul{0.75} & 0.64 & \ul{0.83} & \ul{0.67} & \ul{0.63} & \ul{0.53} & 0.56 & 0.46 & \ul{0.82} & \ul{0.66} & 0.59 & 0.50 \\
    
    \midrule
    \textbf{\name} & \textbf{0.85} & \textbf{0.73} & \textbf{0.85} & \textbf{0.72} & \textbf{0.83} & \textbf{0.75} & \textbf{0.85} & \textbf{0.70} & \textbf{0.75} & \textbf{0.68} & \textbf{0.83} & \textbf{0.75} & \textbf{0.84} & \textbf{0.76} & \ul{0.73} & \ul{0.65}  \\
    \bottomrule
    \end{tabular}
    }
    \label{tab:slinkpred_trind}
\end{table*}

\subsubsection{Dynamic Link Signed Weight Prediction} In this section, we test the efficacy of the methods in predicting the signed weights of links at $t$ given the graph until $T_B^{-1}(t)$. Table~\ref{tab:linkswtpred} shows the performance of different methods on \btcotc and \btcalpha datasets (as only these have weighted edges). \name outperforms or matches the baselines in getting a high \rsq and low \rmse and \kldiv scores. While \name consistently achieves a low \kldiv, there isn't a significant improvement in \rmse and \rsq measures. In particular, the close-to-zero \rsq values indicate that all the models fail to learn the variability of the weights around its mean. This could be attributed to the lack of node features in these datasets or/and implicit conditional independence assumption of the weights given neighborhood features in \gnns~\cite{jia2020residual}. Our results leave room for future research to improve upon our model in predicting signed weights of future links in dynamic signed networks.

\subsection{Transductive and Inductive Settings}
In this section, we compare the performance of the methods in transductive and inductive settings. A \textit{transductive} setting is one where all the links in the test set have nodes that were seen earlier during the training phase. On the other hand, in an \textit{inductive} setting, all the links in the test set have nodes that were not seen earlier during training. We sample the original test set to form a transductive and an inductive test set for each dataset and test our previously trained models on each of these sets. Table~\ref{tab:slinkpred_trind} shows the performance on the \textit{dynamic link signed existence prediction} on these two settings. We find that \name outperforms the baselines in all but two cases with an improvement of up to $16\%$ in \fonewt and $22\%$ in \fonemac for the transductive setting and $17\%$ and $21\%$, respectively, for the inductive setting. \tgn is the closest baseline on average but interestingly, we find that it is often outperformed by static signed methods (\sgcn and \sigat) specifically in the inductive setting. In particular, we note that \sgcn beats \name in the inductive setting of \epinions by a margin of $15\%$ and $17\%$ in \fonewt and \fonemac values respectively. One reason for this observation could be the sparsity of the inductive nodes in the \epinions dataset. Around $75\%$ of the inductive nodes have a degree equal to $1$, i.e., they interact only once. \name is designed specifically to model nodes from their interactions. Node features are used only in the final propagation step and may not affect the node embeddings too much. Future research can study more powerful mechanisms for learning node embeddings from memories to handle highly inductive cases. 

\subsection{Ablation Study}
In this section, we conduct an ablation study to examine \name's sensitivity to each component. In particular, we consider three variants of \name. \textbf{(1) - BA}, which removes the balanced aggregation (BA) of memories. This is equivalent to the original \tgn model. \textbf{(2) - \emb}, where we remove the long-term propagation using a Graph Transformer to form embeddings (i.e., Step 4). The concatenated memories are used as embeddings. \textbf{(3) - \mem}, where we remove the memories and message generation steps. We directly propagate the node features through the aggregated graph until that time via a transformer to find embeddings. 

Table~\ref{tab:ablation} shows the ablation results for the $3$-class dynamic link signed existence prediction. We find that each component is essential for the superior performance of \name as it consistently outperforms its ablation counterparts across datasets. Our results further show that memories are the most important component of the architecture, followed by long-term embedding propagation, and finally balanced aggregation. In fact, removing balanced aggregation can result in a performance comparable to removing the long-term embedding propagation (see \epinions). Thus, balanced aggregation is an essential advance over the original \tgn model for representation learning for dynamic signed networks.

\begin{table}[tb]
    \centering
    \caption{Ablation study of \name on the multiclass dynamic link signed existence prediction. Higher \fonewt and \fonemac are desired. Bold values show the best performing and underlined values show the second best method. 
    }
    \vspace{-0.2cm}
    \renewcommand{\arraystretch}{1}
    \resizebox{0.49\textwidth}{!} {
    \begin{tabular}{c | c c c c c c c c}
    \toprule
    \multirow{2}{*}{\textbf{Method}} & \multicolumn{2}{c}{\textbf{\btcotc}} & \multicolumn{2}{c}{\textbf{\btcalpha}} & \multicolumn{2}{c}{\textbf{\wikirfa}} & \multicolumn{2}{c}{\textbf{\epinions}} \\
    & \fonewt & \fonemac & \fonewt & \fonemac & \fonewt & \fonemac & \fonewt & \fonemac \\
    \midrule
    \textbf{\name} & \textbf{0.88} & \textbf{0.76} & \textbf{0.86} & \textbf{0.73} & \textbf{0.91} & \textbf{0.80} & \textbf{0.82} & \textbf{0.74} \\
    
    \multicolumn{1}{l|}{- \ba} & \textbf{0.88} & \textbf{0.76} & \textbf{0.86} & \ul{0.72} & \ul{0.85} & \ul{0.71} & \ul{0.79} & 0.68 \\
    
    \multicolumn{1}{l|}{- \emb} & 0.71 & 0.57 & 0.72 & 0.59 & 0.79 & 0.67 & 0.78 & \ul{0.71} \\

    \multicolumn{1}{l|}{- \mem} & 0.67 & 0.54 & 0.71 & 0.57 & 0.63 & 0.46 & 0.70 & 0.54 \\
    
    \bottomrule
    \end{tabular}
    }
    \label{tab:ablation}
\end{table}

\subsection{Efficiency}
We compare the efficiency of \name against baselines in terms of training time and memory usage. This is quantified by the per epoch training time in seconds and the allocated GPU memory in bytes for the largest \epinions dataset. First, we note that both \sigat and \sgclstm fail to run on this dataset as they ran out of GPU memory while allocating space during training. \sgcn takes $11.9$ s on average to train one epoch while allocating $862$ MB. On the other hand, more sophisticated \tgn and \name take more time and space resources to train. In particular, \tgn takes $151.4$ s per epoch and requires $11.9$ GB, while \name takes $335.2$ s per epoch with $12.9$ GB of GPU memory. Thus, the superior performance of \name comes at a cost of more training time and GPU resources.

\section{Conclusion}

Our work provides a novel framework to learn and test representations of dynamic signed networks. Our proposed model \name is shown to outperform the baselines across different prediction tasks over dynamic links.
We hope that our model opens new future avenues to study signed dynamic \gnns.
Our method aggregates only first-order signed information from the neighborhood. Recently proposed propagation-based methods such as APAN \cite{wang2021apan} can be extended to encode signed information from the k-hop neighborhood. Other extensions can focus on enforcing other theories of signed network configuration such as status theory \cite{ficsek1991participation,leskovec2010signed}. 
Another drawback of our model is its incapability in predicting the weight distributions. Regression tasks on graphs are known to be notorious and are largely understudied for dynamic graphs. Future works can develop effective models in this vein to learn the entire distribution of signed ratings.




\balance 

\bibliographystyle{ACM-Reference-Format}
\bibliography{citations}


\end{document}